\documentclass{article}

\PassOptionsToPackage{colorlinks=true, citecolor=blue, linkcolor=blue, urlcolor=black}{hyperref}

\usepackage{arxiv}
\usepackage[utf8]{inputenc}
\usepackage[T1]{fontenc}
\usepackage{url}
\usepackage{booktabs}
\usepackage{nicefrac}
\usepackage{microtype}
\usepackage{lipsum}
\usepackage{graphicx}
\usepackage{natbib}
\usepackage{doi}
\usepackage{algorithm}
\usepackage{algorithmicx}
\usepackage{multirow}
\usepackage{multicol}
\usepackage{amsmath,amssymb,amsfonts}
\usepackage{amsthm}
\usepackage{mathrsfs}
\usepackage{xcolor}
\usepackage{float}

\title{Self-supervised Surface-related Multiple Suppression with Multidimensional Convolution}

\author{ \href{https://orcid.org/0000-0001-8868-7967}{\includegraphics[scale=0.06]{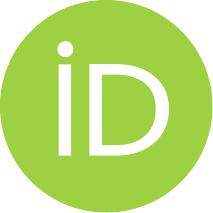}\hspace{1mm}Shijun~Cheng}\\
	Division of Physical Science and Engineering\\
	King Abdullah University of Science and Technology\\
	Thuwal 23955-6900, Saudi Arabia \\
	\texttt{sjcheng.academic@gmail.com} \\
        \And
	\href{https://orcid.org/0009-0004-2532-4615}{\includegraphics[scale=0.06]{orcid.pdf}\hspace{1mm}Ning Wang} \\
	Division of Physical Science and Engineering\\
	King Abdullah University of Science and Technology\\
	Thuwal 23955-6900, Saudi Arabia \\
	\texttt{ning.wang.2@kaust.edu.sa} \\
        \And
	\href{https://orcid.org/0000-0002-9363-9799}{\includegraphics[scale=0.06]{orcid.pdf}\hspace{1mm}Tariq~Alkhalifah} \\
	Division of Physical Science and Engineering\\
	King Abdullah University of Science and Technology\\
	Thuwal 23955-6900, Saudi Arabia \\
	\texttt{tariq.alkhalifah@kaust.edu.sa} \\
}

\renewcommand{\shorttitle}{Self-supervised surface-related multiple suppression}

\hypersetup{
pdftitle={A template for the arxiv style},
pdfsubject={q-bio.NC, q-bio.QM},
pdfauthor={David S.~Hippocampus, Elias D.~Striatum},
pdfkeywords={First keyword, Second keyword, More},
}

\begin{document}
\maketitle

\begin{abstract}
Surface-related multiples pose significant challenges in seismic data processing, often obscuring primary reflections and reducing imaging quality. Traditional methods rely on computationally expensive algorithms, the prior knowledge of subsurface model, or accurate wavelet estimation, while supervised learning approaches require clean labels, which are impractical for real data. Thus, we propose a self-supervised learning framework for surface-related multiple suppression, leveraging multi-dimensional convolution to generate multiples from the observed data and a two-stage training strategy comprising a warm-up and an iterative data refinement stage, so the network learns to remove the multiples. The framework eliminates the need for labeled data by iteratively refining predictions using multiples augmented inputs and pseudo-labels. Numerical examples demonstrate that the proposed method effectively suppresses surface-related multiples while preserving primary reflections. Migration results confirm its ability to reduce artifacts and improve imaging quality.
\end{abstract}


\section{\textbf{Introduction}}
Real seismic data often suffer from the presence of surface-related multiples, which are a type of coherent noise caused by seismic waves reflecting multiple times between the free surface and subsurface interfaces. These multiples can overwhelm primary reflections, reduce the clarity of subsurface images, and complicate subsequent data processing steps such as velocity analysis and inversion. Therefore, effectively suppressing free-surface multiples is crucial for enhancing seismic imaging quality and improving the accuracy of subsurface structural interpretation \citep{verschuur2013seismic}. However, multiple suppression remains a challenging task due to the overlapping nature of multiples with primary reflections in both time and frequency domains, as well as their relatively stronger signature especially later in the recording time. 

Traditional methods for surface-related multiples suppression have been widely applied in seismic processing workflows. Generally, it can be categorized into two classes: filtering-based methods and wave-equation-based methods. The filtering-based approaches exploit differences in seismic event characteristics, such as apparent velocity and frequency content, to separate multiples from primary reflections \citep{hampson1986inverse, hu2008robust}. Among these, the Radon transform-based technique is particularly popular due to its effectiveness in distinguishing multiples from primaries in the time–slowness ($\tau–p$) domain, facilitating robust multiple attenuation through selective filtering \citep{foster1992suppression, nowak2006amplitude, jiang2020adaptive, song2022multiple}. On the other hand, wave-equation-based methods explicitly use physical wave propagation mechanisms to predict multiples \citep{hokstad20063d, herrmann2008adaptive, van2009estimating, van2009estimation, dragoset2010perspective, zuberi2013imaging, lopez2015closed, he2022elimination, sui2024seismic}, where surface-related multiple elimination (SRME) being the most representative example \citep{verschuur1992adaptive, berkhout1997estimation, verschuur1997estimation}. SRME employs a data-driven convolutional process to predict multiples directly from recorded seismic data without requiring prior knowledge of the subsurface velocity structure or reflectivity. This prediction estimating the source wavelet and using it to apply a normalizing multi-dimensional convolution (MDC) of the seismic data with themselves in both time and space, effectively reconstructing the multiple reflections originating from the free surface. Then, adaptive subtraction is used to eliminate the multiples.

While aforementioned approaches are theoretically sound and effective in many scenarios, they suffer from several limitations. Filtering-based methods, although efficient, often suffer from insufficient resolution in complex geological settings and would attenuate some of the primary reflections energy due to their overlapping characteristics with the multiples in the transformed domain. Wave-equation-based approaches, while theoretically accurate, are computationally expensive, often requiring substantial processing time and computational resources. Meanwhile, their performance heavily relies on the prior knowledge of the subsurface models or accurate wavelet estimations \citep{weglein1999multiple}. 

To address these limitations, recent studies have explored the use of neural networks (NNs) for multiple suppression tasks \citep{mousavi2024applications}. While these NN-based approaches have shown promise, most of them are built upon supervised learning (SL) frameworks, where the training process depends on paired datasets with clean (multiple-free) and noisy (multiple-contaminated) data \citep{liu2021adaptive, ma2024u}. For example, \cite{siahkoohi2019surface} trained a convolutional neural network (CNN) in an SL manner to approximate the nonlinear relationship between seismic data containing surface-related multiples and the primaries, where the primaries were obtained using a sparse-inversion-based algorithm. \cite{li2021adaptive} applied a U-Net-based network to perform adaptive subtraction as a nonlinear problem, significantly improving multiple removal compared to traditional linear regression method. \cite{wang2022seismic} proposed a CNN with an encoding-decoding architecture, trained through data augmentation, to reconstruct primaries and robustly suppress multiples and background noise from marine seismic data. These studies demonstrated the initial potential of applying NNs to tackle the task of multiple suppression. 

However, most SL-based methods are trained on synthetic datasets, where clean and noisy pairs can be artificially generated. There exists a significant distributional gap between synthetic and real data, which often leads to poor generalization performance when these models are applied to field data \citep{alkhalifah2022mlreal}. Thus, SL-based approaches for multiple suppression face a fundamental challenge: obtaining clean labels for real seismic data is practically impossible because the true multiple-free wavefield is unknown. Using labels from the application of conventional surface multiple elimination methods is both costly and will limit the trained neural network performance to the quality of these labels. As an alternative to SL, self-supervised learning (SSL) methods naturally overcome generalization issues by eliminating the need for labeled data, thus gradually attracting increasing attention in seismic processing tasks \citep{saad2020deep, birnie2021potential, liu2024gabor, cheng2024self, cheng2025self, li2024robust, li2025unsupervised}. 

In the realm of multiple suppression, several studies have proposed training NNs using SSL approaches to improve performance on field data.
\cite{wang2022unsupervised2, wang2023surface} used an ensemble of three SSL NNs to refine the surface-related multiple predictions produced by the traditional SRME, which adjust amplitude and phase differences to improve multiple suppression and preserve the primaries. However, the quality of their multiple suppression heavily depends on the accuracy of the predicted surface-related multiples provided by SRME. SRME itself requires extensive iterations and relies on prior information that is often difficult to obtain accurately, which can compromise efficiency and overall performance. Similarly, \cite{wang2022unsupervised1, wang2023unsupervised} employed adaptive virtual events to predict internal multiples and trained multiple SSL networks with joint constraints to map these predictions to true multiples. However, one limitation is that the method relies on well logging data or expert judgment to estimate the number of expected subsurface scattering interfaces, and the iterative suppression for each interface is time-consuming. \cite{li2023unsupervised} proposed unrolling the iterative steps of the fast iterative shrinkage thresholding algorithm into a network and replaces the traditional thresholding operation with a U-Net to adaptively suppress multiples in an SSL framework. Yet, they requires convolution operations at every iteration, leading to high computational cost and limiting the effective suppression of multiples in large datasets. \cite{liu2024physics} used virtual events to generate initial internal multiples as prior information and employs a physics-informed SSL network with a hybrid loss (combining mean absolute error and phase resemblance) to adaptively correct amplitude and phase distortions. However, the performance of their method depends on having relatively accurate initial internal multiples as prior information, which is difficult to ensure, especially with complex real data.

In this paper, we propose a novel SSL method for surface-related multiple suppression that is free of the above mentioned challenges. Unlike existing approaches, our method does not require any prior information such as wavelets or subsurface velocity models, and it only requires a MDC to be preformed once to generate surface-related multiples. These generated multiples share the essential spatial and temporal features of real surface multiples, but differ in wavelength and amplitude, enabling us to avoid repeatedly performing MDC while still capturing the necessary characteristics of the multiples. In our SSL framework, the network training is divided into a warm-up stage and an iterative data refinement (IDR) stage. By iteratively refining model predictions, we effectively avoids the dependency on clean labels while significantly enhancing efficiency, robustness, and generalization on field data. We will share two synthetic examples and one marine field data to demonstrate the effectiveness of our approach.

\section{\textbf{Method}}
This section introduces our SSL framework for surface-related multiple suppression, which consists of three main components. First, we employ multi-dimensional convolution (MDC) to generate synthetic surface-related multiples from the observed data. Second, inspired by the Noisier2Noise approach, we construct noisier (multiples augmented) inputs and less noisy pseudo-labels to initiate the training of our network. Finally, we refine the model in two sequential stages, namely warm-up and IDR, to progressively enhance the suppression of surface-related multiples.

\subsection{Multi-dimensional convolution}
Surface-related multiples exhibit distinct spatial and temporal characteristics, making them predictable to some extent through convolution-based operations. In this study, we consider a medium where sources and receivers co-located at the free surface. Under these conditions, artificial surface-related multiples, denoted as ${d}_{\text{m}}$, can be  constructed by applying MDC to the raw seismic data ${d}_{\text{raw}}$, free of direct arrivals. This procedure yields a data representation that preserves the kinematic and dynamic properties of the true surface-related multiples, with an additional source wavelet signature courtesy of the convolution process. As such, ${d}_{\text{m}}$ is treated as a synthetic approximation of the surface-related multiples extracted from the observed data. The process can be expressed mathematically as:
\begin{equation}\label{eq1}
{d}_{\text{m}}(\textbf{x}_{R},\textbf{x}_{S}) = \int _{\Lambda _{R}}{d}_{\text{raw}}(\textbf{x}_{R},\textbf{x}_{R}'){d}_{\text{raw}}(\textbf{x}_{R}',\textbf{x}_{S})d\textbf{x}_{R}'
,
\end{equation}
where $\textbf{x}_{S}$, $\textbf{x}_{R}$, and $\textbf{x}_{R}'$ represent the sources, receivers, and virtual sources, respectively. \( \Lambda _{R} \) represents the acquisition surface. ${d}_{\text{raw}}$
are raw seismic data containing surface-related multiples.  ${d}_{\text{m}}$ represents generated artificial surface-related multiples, which serves as a key component in our SSL framework, ensuring the network can learn to suppress surface-related multiples effectively without relying on manually labeled clean data.

\subsection{The Noisier2Noise concept}
Our method is inspired by the Noisier2Noise concept \citep{moran2020noisier2noise}, which is an SSL denoising strategy that uses noisy data as both the input and pseudo-labels. Specifically, one creates an even noisier version of the data by adding artificial noise, while the relatively less noisy original data serve as pseudo-labels. This paradigm has been shown to be effective in attenuating many types of seismic noise \citep{cheng2024effective}. 

In our context, the raw seismic data $d_{\text{raw}}$ already contain surface-related multiples. We generate additional synthetic multiples $d_\text{m}$ using MDC and add them (optionally with a scaling factor) to $d_\text{raw}$ to obtain a version of the data with stronger multiple energy. This noisier data are fed into the network, while the original $d_\text{raw}$ acts as the less noisy pseudo-label. By mapping from these intensively contaminated data to the relatively less contaminated (raw) data, the network learns an initial multiple suppression capability without any requirement for clean labels.

\subsection{Self-supervised surface-related multiple suppression}
Our SSL training proceeds in two main stages: a warm-up phase followed by an IDR phase. Each stage serves a distinct purpose, progressively enhancing the network's ability to suppress surface-related multiples. 

The goal of the warm-up stage is to initialize the network with a rough understanding of how to suppress surface-related multiples. In this stage, we construct the network input by adding the MDC-generated multiples $d_\text{m}$ to the raw seismic data $d_\text{raw}$. The original data $d_\text{raw}$ serve as the pseudo-labels. On the other hand, the input to the network is given by
\begin{equation}\label{eq2}
    d_\text{input} = d_\text{raw} + \alpha d_\text{m},
\end{equation}
where $\alpha$ is a scaling factor that adjusts the energy of the synthetic multiples. This setup follows the Noisier2Noise paradigm: the network learns an initial mapping from noisier input (stronger multiples) to less noisy targets (raw data), establishing a baseline understanding of multiple suppression. 

After the warm-up stage, we further refine the model through an iterative SSL process. This stage builds on the pre-trained weights from the warm-up stage and operates in an epoch-based refinement cycle. In each training epoch $k$, the network produces an output $\hat{d}^{(k)}$, which becomes the pseudo-label for the next epoch. To build the input for epoch $(k+1)$, we similarly add the MDC-generated multiples to the output of the previous epoch:
\begin{equation}\label{eq3}
    d_\text{input}^{(k+1)} = \hat{d}^{(k)} + \alpha d_\text{m}.
\end{equation}
Through this iterative refinement, the network progressively improves its ability to distinguish and suppress surface-related multiples, while reducing reliance on pseudo-label inaccuracies from earlier epochs. This happens because in every epoch, starting from the warm up, the network experiences cleaner pseudo labels courtesy of it learning to remove more noise. In other words, as we progress the input and labels for training the network become progressively cleaner until $\hat{d}^{(k)}$ becomes approximately the clean data, and the network learns to remove $\alpha d_m$. 

In both the warm-up and IDR stages, the scaling factor $\alpha$ plays a key role in matching the amplitude of the MDC-generated multiples to the real multiple energy in the raw data. Ideally, $\alpha$ should scale the MDC-generated multiples’ energy to match the observed surface-related multiples in $d_\text{raw}$. If $\alpha$ is set too large, the network may over-suppress primary reflections of interest along with the multiples. If $\alpha$ is too small, the network may learn only limited suppression. In practice, a robust strategy is to let $\alpha$ vary randomly within a small range around an empirically determined median value. This median can be estimated by visually or quantitatively comparing the amplitude of $\alpha d_\text{m}$ against the observed multiples in $d_\text{raw}$. Once identified, the same range of $\alpha$ values can be shared between the warm-up and IDR stages. 

Since we only perform MDC once at the beginning to generate $d_\text{m}$, our framework avoids repeatedly performing computationally expensive convolution operations. Moreover, this method does not rely on prior information such as wavelets or subsurface velocity models, enhancing both efficiency and applicability to field data. 

\begin{figure}[htbp]
\centering
\includegraphics[width=1\textwidth]{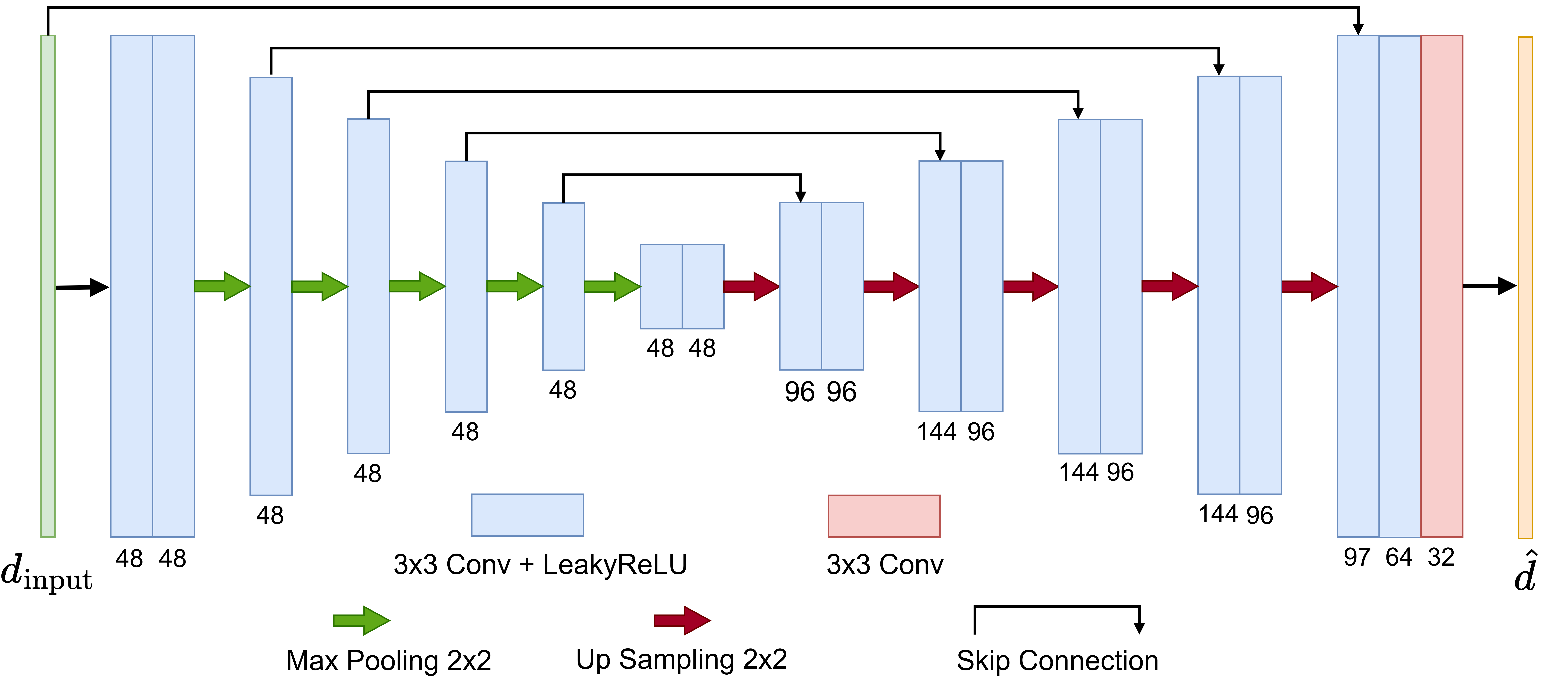}
\caption{The neural network architecture used in our study, where the input $d_\text{input}$ is the seismic data contaminated with surface-related multiples, and the output $\hat{d}$ is the de-multiple result.}
\label{network}
\end{figure}

\subsection{Network Architecture}
We use a modified U-Net as our baseline network for surface-related multiple suppression. We avoid more advanced architectures in order to demonstrate the general applicability of our approach. Figure~\ref{network} illustrates the network architecture, which follows a U-Net-like design but includes specific modifications. The model takes an input shot gather $d_\text{input}$ and produces a prediction $\hat{d}$. The overall structure comprises an encoder and a decoder, with skip connections bridging the corresponding layers. 

Each light-blue block in both the encoder and decoder represents a 3$\times$3 convolution followed by a leaky rectified linear unit (LeakyReLU) activation. The specific number of feature channels is noted beneath each block in the figure. In the encoder path, each stage concludes with a 2$\times$2 max-pooling operation (green arrow), reducing the spatial resolution by half while preserving critical seismic information. The decoder path begins each stage with a 2$\times$2 up-sampling operation (red arrow), restoring the spatial dimensions. Skip connections (black arrows) then reintroduce high-resolution details from the encoder, aiding the network in distinguishing primary reflections from multiple energy. Finally, a 3$\times$3 convolution layer (pink block) merges the feature maps into the output shot gather $\hat{d}$. 
\section{\textbf{Synthetic examples}}
We here first evaluate our method through two synthetic examples, which are performed on two different models: a layered model and the Otway model. Both models use a uniform grid spacing of 5 m. We simulate the seismic wavefield with an acoustic propagator using DeepWave \citep{richardson_alan_2023}. Both models assumes constant density, and the motion is excited by a Ricker wavelet with a peak frequency of 10 Hz. To generate the synthetic multiple data, we employ an MDC operator implemented in PyLops \citep{ravasi2020pylops}. We emphasis that before performing the MDC operation, we remove direct arrivals from the recorded data. This ensures that the MDC-generated multiples predominantly capture surface-related multiple energy, free from strong direct-wave contamination.

Throughout both synthetic examples and the following field application, we use a combined loss function that includes mean absolute error and multi-scale structural similarity. For optimization, we adopt the AdamW optimizer \citep{loshchilov2017decoupled} and conduct all training and inference on an RTX 8000 GPU with 48 GB of memory.

\begin{figure}[htbp]
\centering
\includegraphics[width=0.5\textwidth]{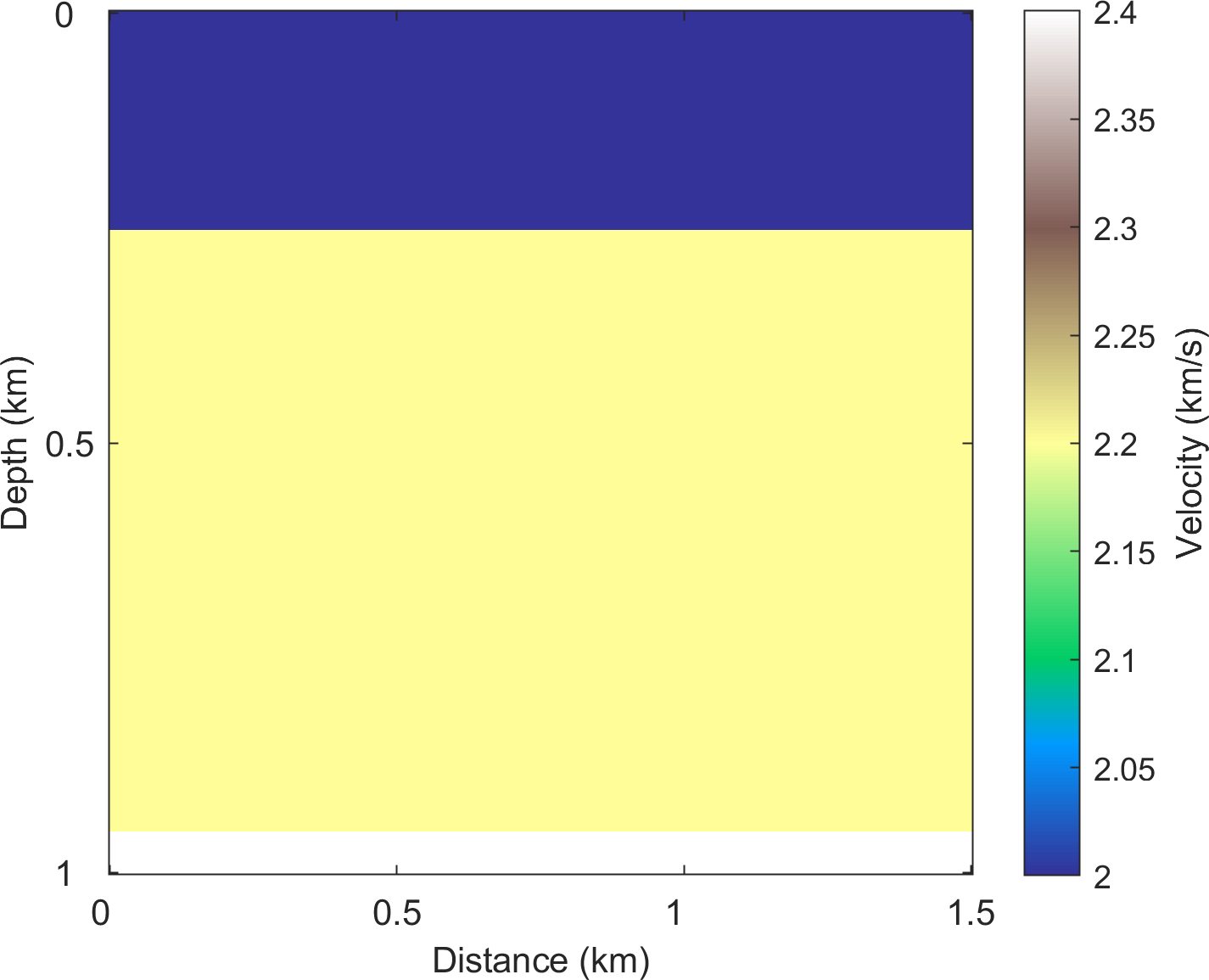}
\caption{The layered velocity model.}
\label{fig1}
\end{figure}

\begin{figure}[htbp]
\centering
\includegraphics[width=0.9\textwidth]{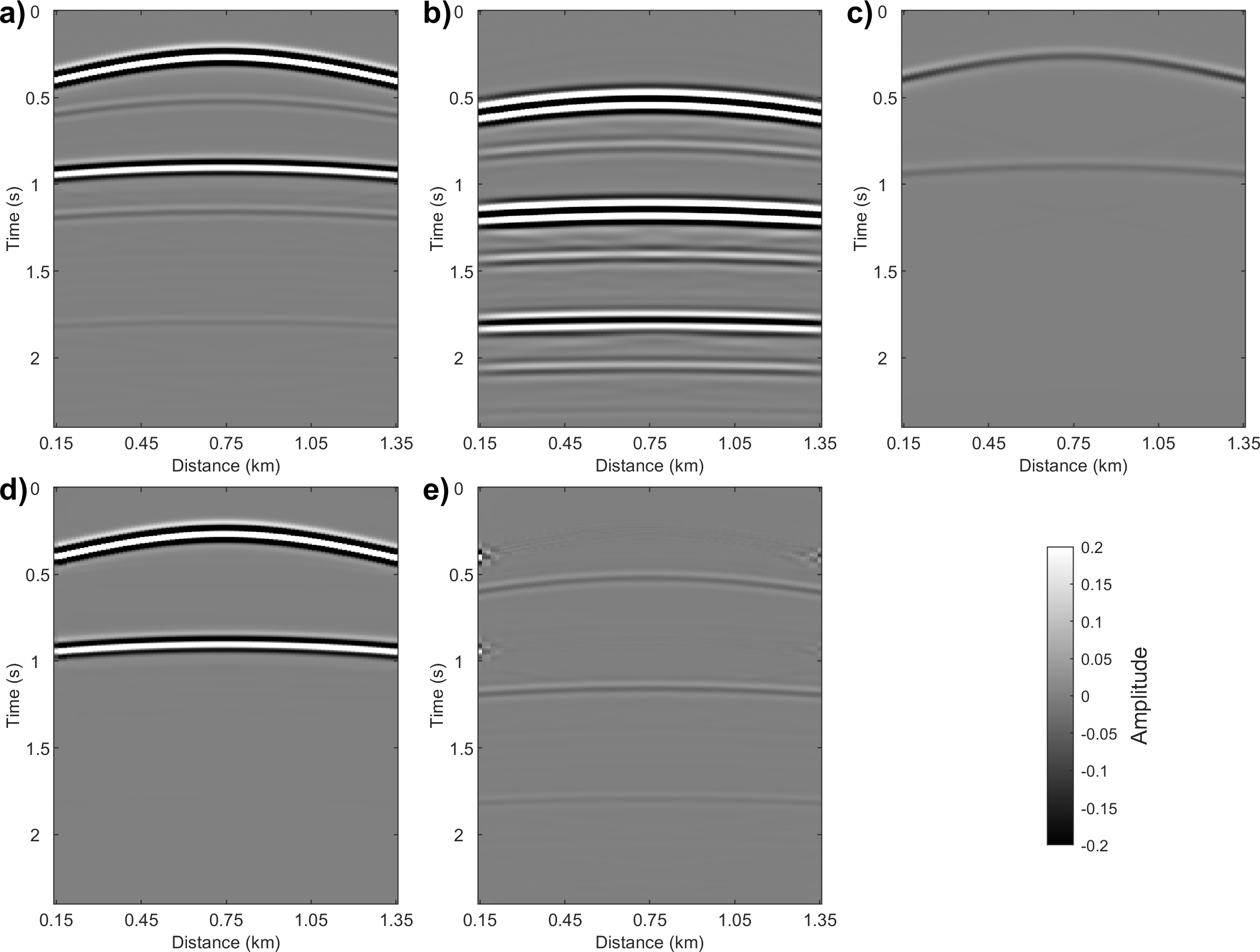}
\caption{(a) A representative shot gather from the layered model using a free-surface boundary. (b) Multiples data generated using the MDC operation from the raw shot gather in (a). (c) A reference shot gather simulated with an absorbing boundary condition. (d) The predicted result for the original data in (a). (e) The residual between the original data (a) and our prediction (d). }
\label{fig2}
\end{figure}

\subsection{Layered model}
We first construct a three-layer model on a $201\times301$ grid. Figure \ref{fig1} illustrates the velocity distribution, where the thicknesses of the three layers (from top to bottom) are 0.25 km, 0.7 km, and 0.05 km, respectively, and their corresponding velocities are 2.0 km/s, 2.2 km/s, and 2.4 km/s. We impose a free-surface boundary at the top of the model to simulate the seismic data contaminated with surface-related multiples, while the remaining boundaries employ convolutional perfectly matched layers (C-PML) to absorb outgoing waves. We simulate 101 shots along the top boundary, each spaced 15 m apart, and placed 101 receivers also spaced 15 m apart at the same depth to record data up to 2.5 s. 

Figure \ref{fig2}a displays a representative shot gather from this model. Figure \ref{fig2}b shows the synthetic surface-related multiples generated by the MDC operator, while Figure \ref{fig2}c presents the corresponding shot gather obtained by replacing the free-surface boundary with a C-PML top boundary (thus removing the prospect of surface-related multiples). From Figures \ref{fig2}a–c, we can see that the MDC-generated multiples share the same spatial and temporal locations as the surface-related multiples in the raw shot gather, but exhibit different amplitudes and wavelengths due to the extra source signature involved in the convolution process. For approaches based on SRME, we divide the output of MDC by the source wavelet. This step, which requires accurate knowledge of the source wavelet, is not need here at network will inherently learn to do that since the label, given by the raw data, and the pseudo labels, do not have this additional wavelet signature. Note that the primaries recorded under a free-surface boundary have different amplitudes compared to those from the absorbing-boundary simulation, yet their phases remain consistent. We will use the absorbing boundary based data as reference to help us identify the primaries.  

For training, we use the 101 shot gathers simulated with a free-surface boundary condition, together with the corresponding MDC-generated multiples. We augment the data by horizontally (left-right) and vertically (up-down) flipping entire shot gathers rather than subdividing them into smaller patches. The training batch size is set to 32, and we train for 160 epochs in total, with 50 epochs for the warm-up stage. For both warm-up and IDR phases, the range of $\alpha$ is set to $[0.01, 0.08]$. The learning rate is fixed at $2\times10^{-4}$. Overall, training requires approximately 10 minutes and 24 seconds. 

Figure \ref{fig2}d shows the network’s prediction for the same shot gather in Figure \ref{fig2}a, and Figure \ref{fig2}e presents the residual. We can see that our method effectively suppresses the surface-related multiples while preserving the primary events. Only minor leakage appears near the model boundaries, but this does not affect the subsequent imaging process. 

\begin{figure}[htbp]
\centering
\includegraphics[width=0.5\textwidth]{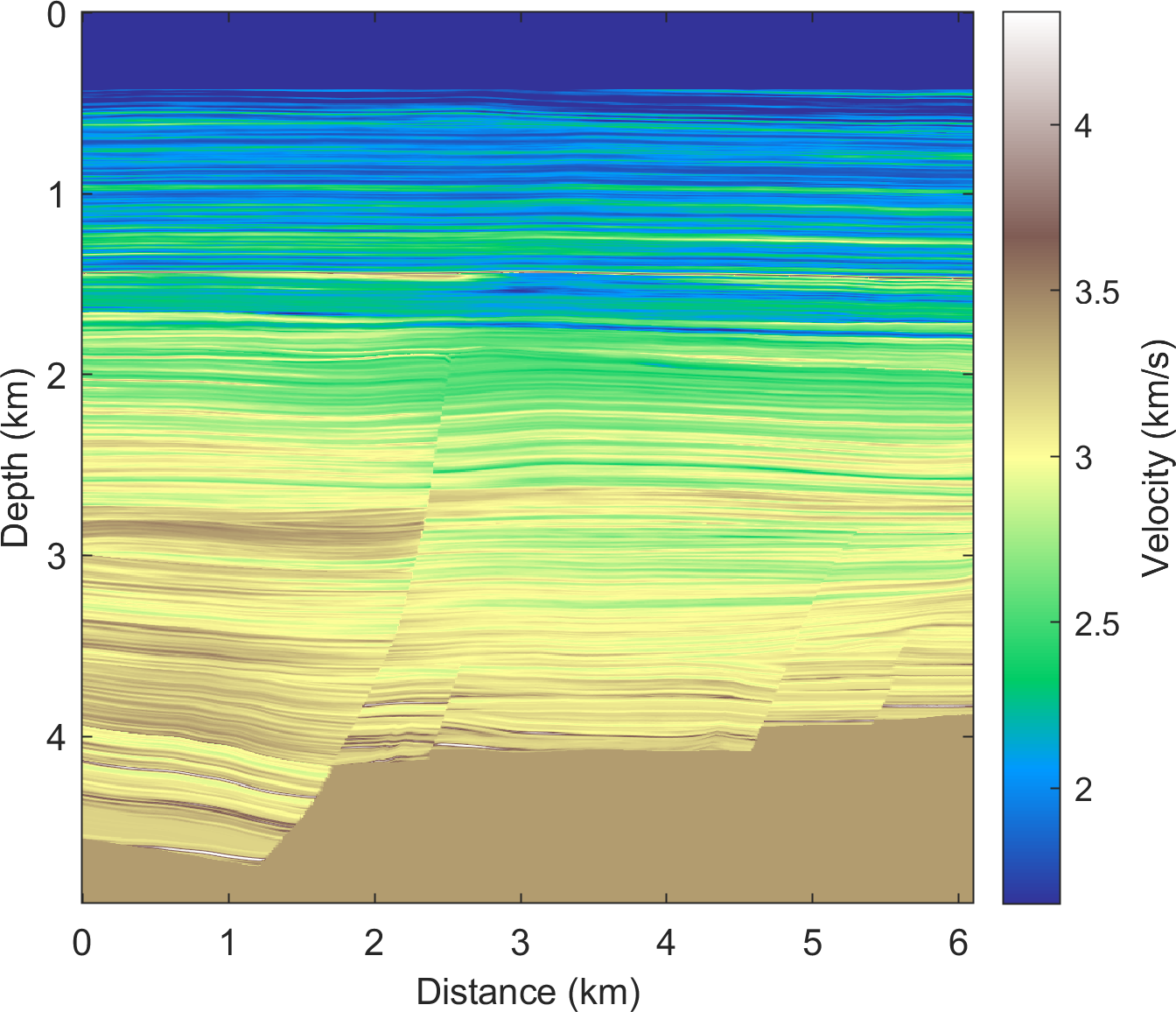}
\caption{The Otway velocity model.}
\label{fig3}
\end{figure}

\subsection{Otway model}
We then evaluate the performance of our proposed method using a more complex Otway model. The velocity model, shown in Figure \ref{fig3}, has original dimensions of \(984 \times 1220\). We apply a free-surface boundary condition and position the first shot at the leftmost side of the surface of the model. Subsequent shots are spaced every three grid points, resulting in a total of 407 shots. For each shot, 407 receivers are evenly distributed along the surface, with one receiver placed every three grid points starting from the leftmost position, and coinciding with the sources locations. 

\begin{figure}[htbp]
\centering
\includegraphics[width=0.9\textwidth]{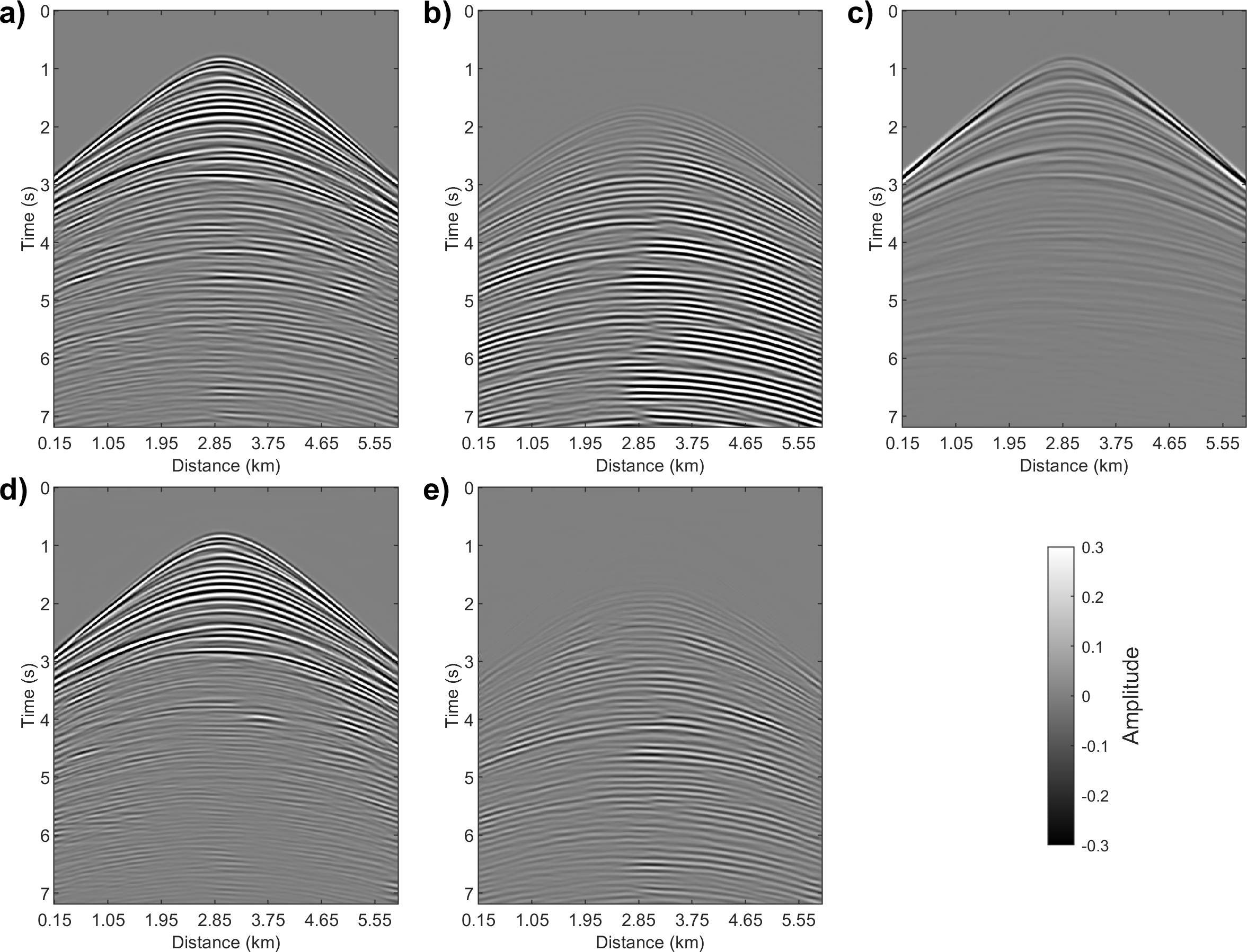}
\caption{Similar with Figure \ref{fig2}, but for the Otway velocity model.}
\label{fig4}
\end{figure}

\begin{figure}[htbp]
\centering
\includegraphics[width=1\textwidth]{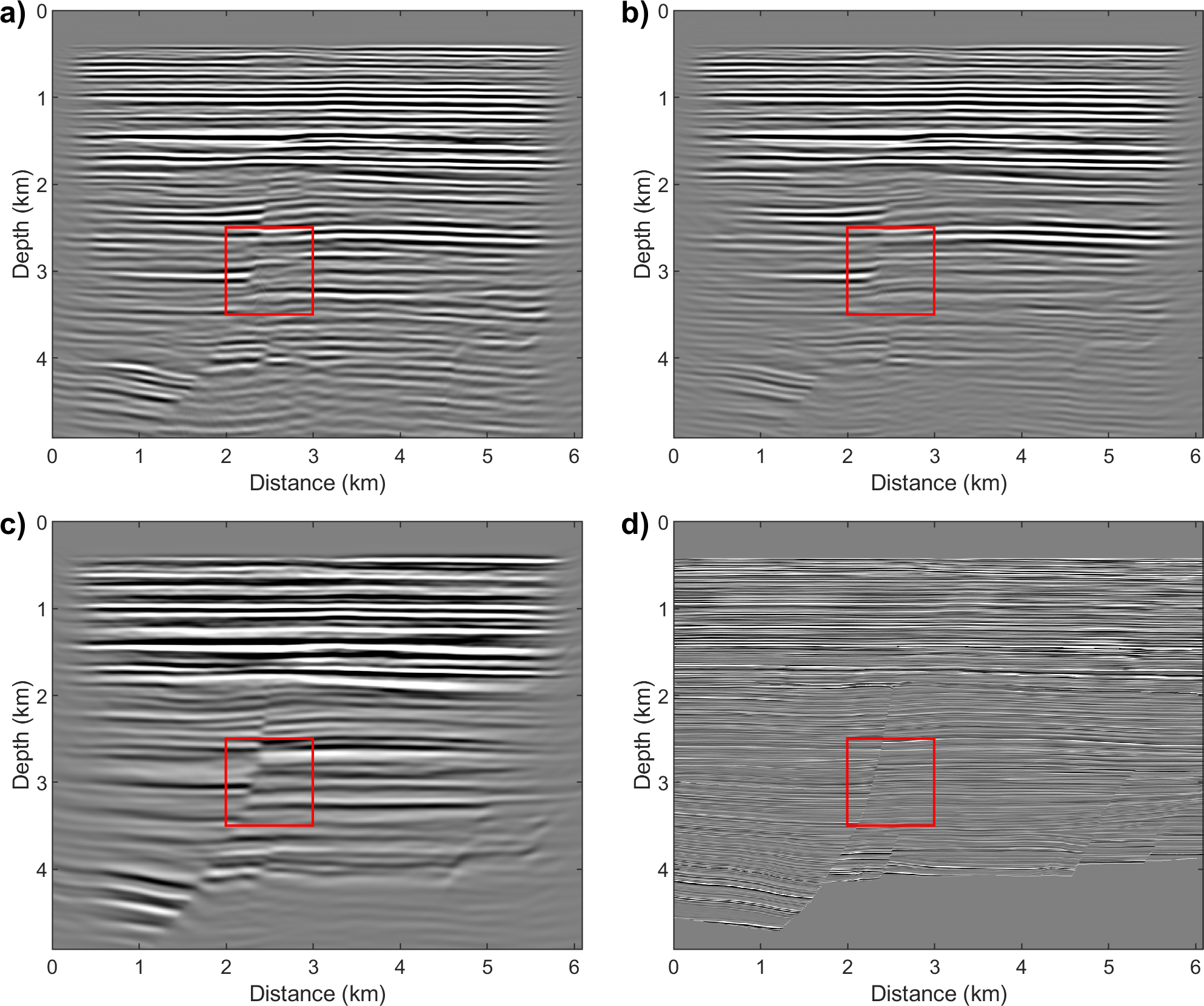}
\caption{Migrated images for the Otway model with the three datasets: (a) The raw data simulated with a free-surface boundary condition. (b) The multiple-suppressed data obtained using the proposed method. (c) The data simulated with an absorbing boundary condition. Panel (d) is the reflectivity model.}
\label{fig5}
\end{figure}

\begin{figure}[htbp]
\centering
\includegraphics[width=1\textwidth]{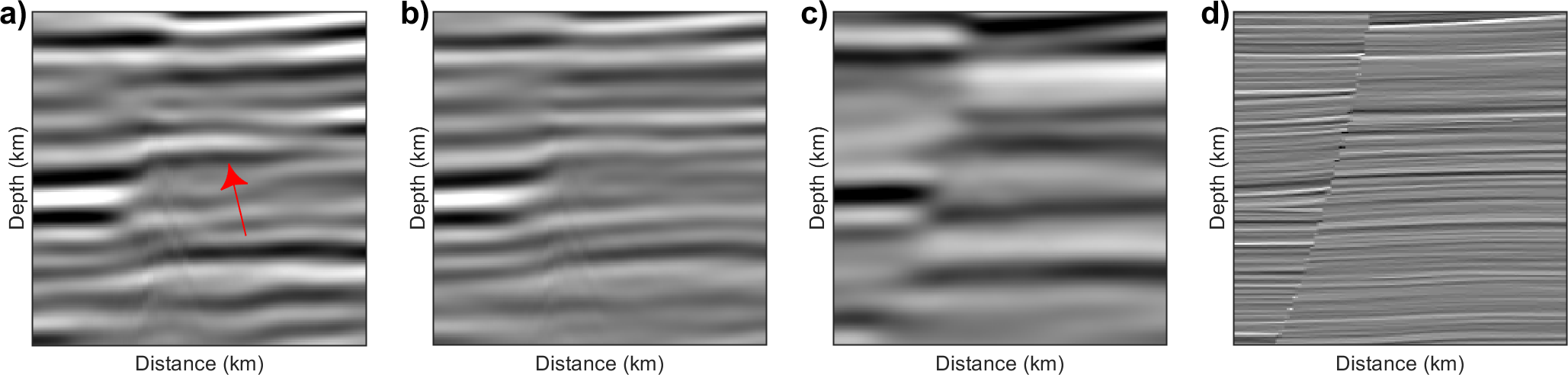}
\caption{Zoomed-in views of the red rectangular region in Figure ~\ref{fig5}, respectively.}
\label{fig6}
\end{figure}

Figure \ref{fig4}a shows a representative shot gather simulated with a free-surface boundary condition, with the direct arrivals removed. We apply the MDC operation to the raw seismic data containing surface-related multiples to generate multiples data. An example of the multiples data constructed using the MDC operation is shown in Figure \ref{fig4}b. For reference, we also simulate a shot gather using an absorbing boundary at the top, shown in Figure~\ref{fig4}c. Consistent with our earlier findings, the MDC-derived multiples occupy the same spatial and temporal regions as the raw data’s surface-related multiples but exhibit distinct amplitudes and wavelengths.

To train our network, we use the 407 shot gathers simulated with a free-surface boundary condition, along with the MDC-generated multiples data. The training batch size is set to 20. The network is trained over 70 epochs, with 50 epochs allocated to the warm-up stage and 20 epochs for the IDR stage. For both warm-up and IDR phases, the range of $\alpha$ is set to $[0.2, 0.8]$. The training process starts with an learning rate of \(2 \times 10^{-4}\), which decays to 0.6 of its value at the 25th and 50th epochs. The training takes approximately 139 minutes. 

Figure \ref{fig4}d presents the network's prediction results for the original shot gather, and Figure \ref{fig4}e shows the residual between the original shot gather and the prediction. By comparing Figures \ref{fig4}e with \ref{fig4}b, we can see that our method effectively suppresses many of the multiples, as evidenced by the matching positions of the residual and the MDC-constructed surface-related multiples. 

To further validate the effectiveness of our approach, we perform imaging using three datasets: the raw data simulated with a free-surface boundary condition, the multiple-suppressed data obtained from our method, and the data simulated with an absorbing boundary condition. The corresponding migration results are shown in Figures \ref{fig5}a, \ref{fig5}b, and \ref{fig5}c, respectively, and panel d represents the reflectivity model. To better highlight the artifacts caused by surface-related multiples, we select the regions marked by red rectangles in Figure \ref{fig5} and provide a zoomed-in view in Figure \ref{fig6}. As indicated by the red arrow in Figure \ref{fig6}a, surface-related multiples cause a fake layer in the migrated image. In contrast, our prediction results successfully suppress the multiples, effectively removing the fake layer from the migration result, as shown in Figure \ref{fig6}b. These findings demonstrate the robustness and effectiveness of our proposed method in suppressing surface-related multiples and improving the quality of seismic imaging. 

\begin{figure}[htbp]
\centering
\includegraphics[width=0.5\textwidth]{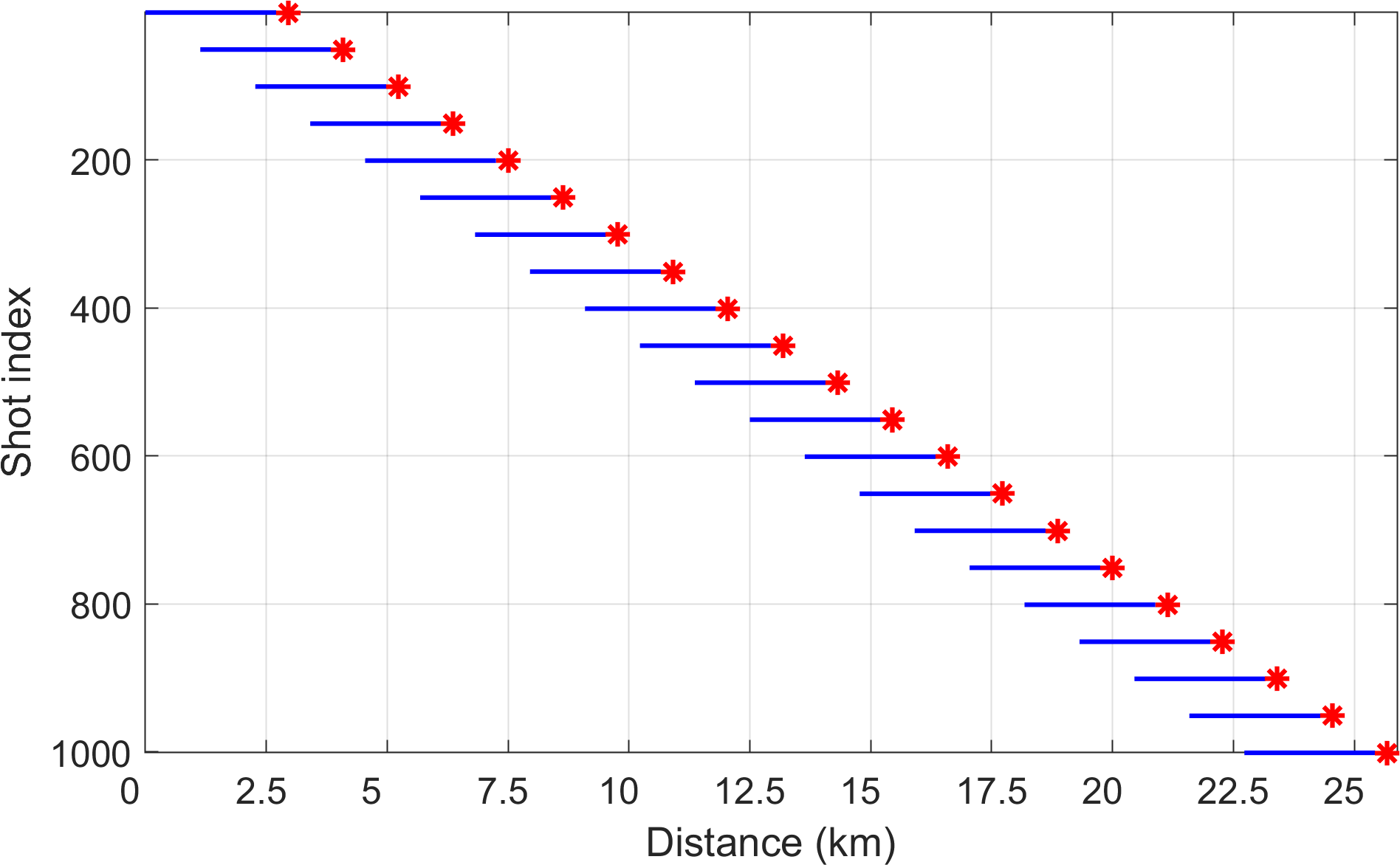}
\caption{Acquisition geometry for the open-source marine dataset, where the blue lines represent the streamers and the red stars represent the sources.}
\label{fig7}
\end{figure}

\begin{figure}[htbp]
\centering
\includegraphics[width=1\textwidth]{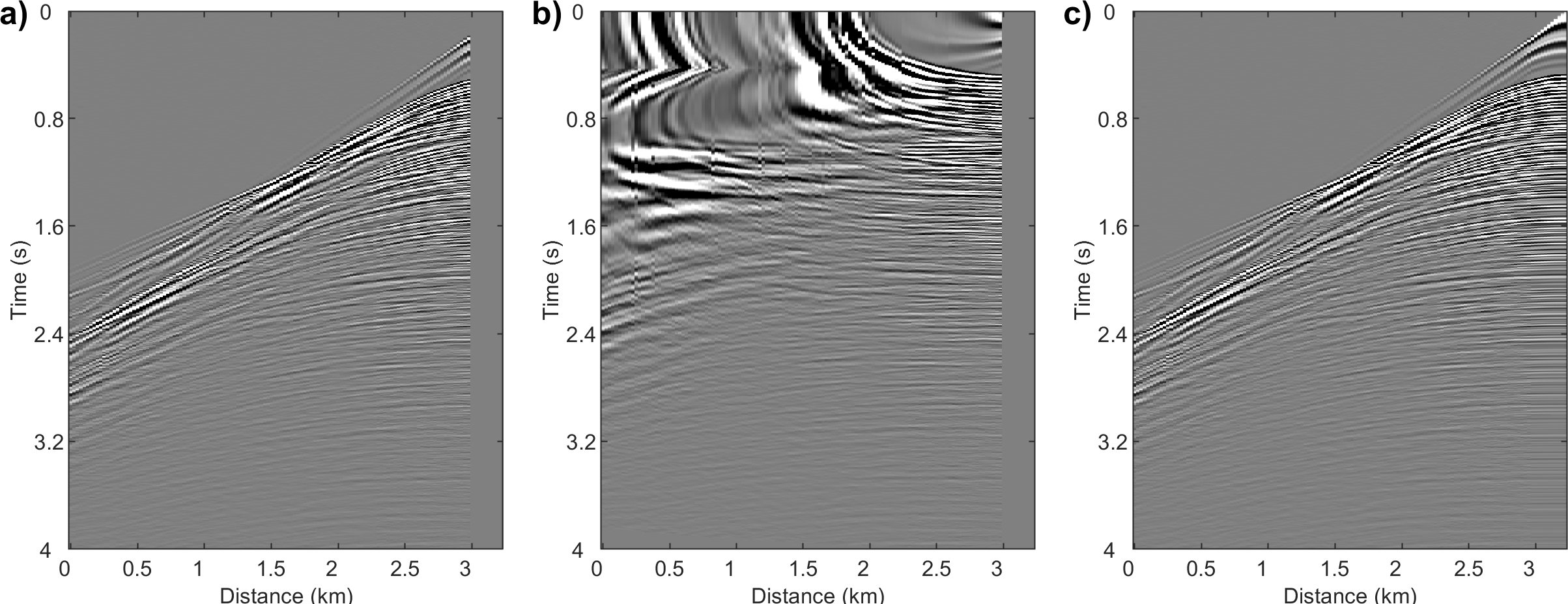}
\caption{(a) A representative shot gather from the marine dataset, with near offsets missing. (b) The gather after a rough NMO correction. (c) The gather after duplicating the nearest recorded traces into the missing near offsets and removing the NMO correction.}
\label{fig8}
\end{figure}

\begin{figure}[htbp]
\centering
\includegraphics[width=0.7\textwidth]{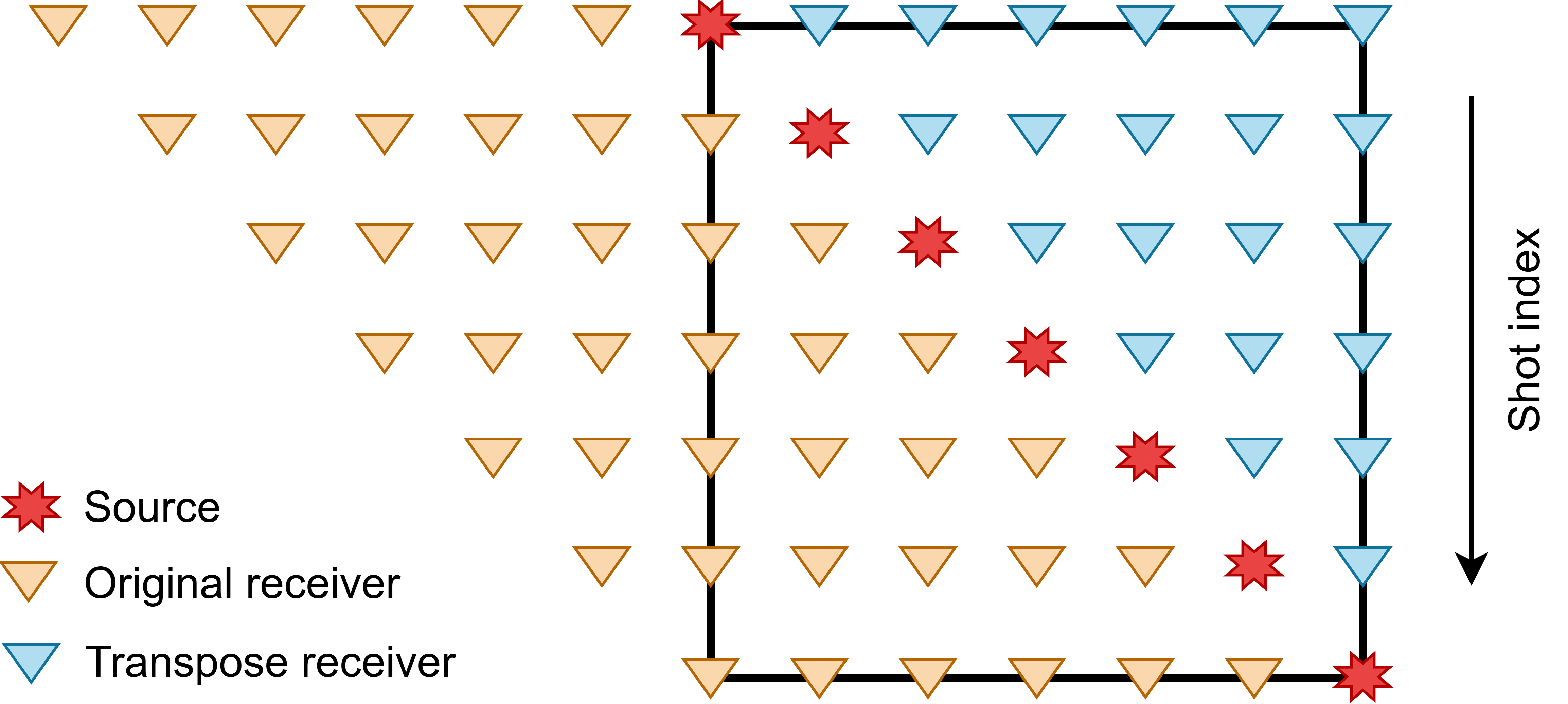}
\caption{As schematic plot demonstrating the shot-receiver reciprocity approach. }
\label{fig9}
\end{figure}

\begin{figure}[htbp]
\centering
\includegraphics[width=1\textwidth]{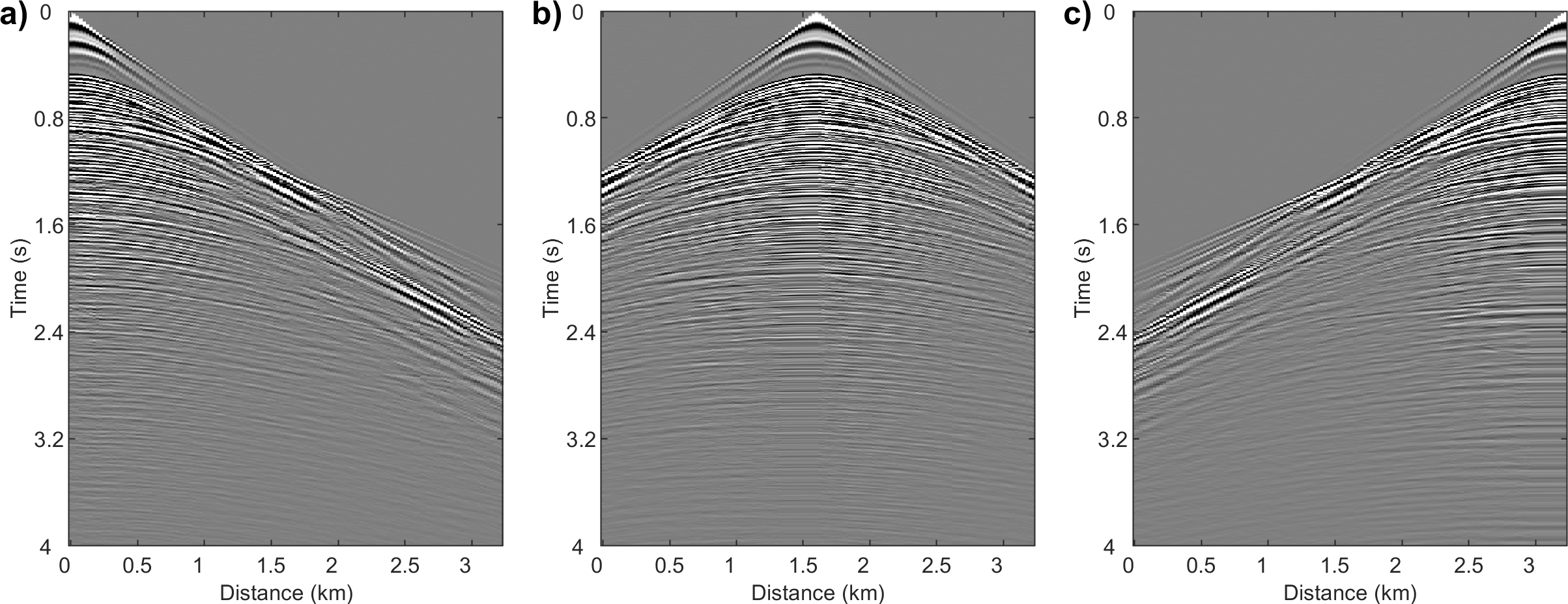}
\caption{Three representative shot gathers after applying near-offset interpolation and shot-receiver reciprocity.}
\label{fig10}
\end{figure}

\begin{figure}[htbp]
\centering
\includegraphics[width=0.8\textwidth]{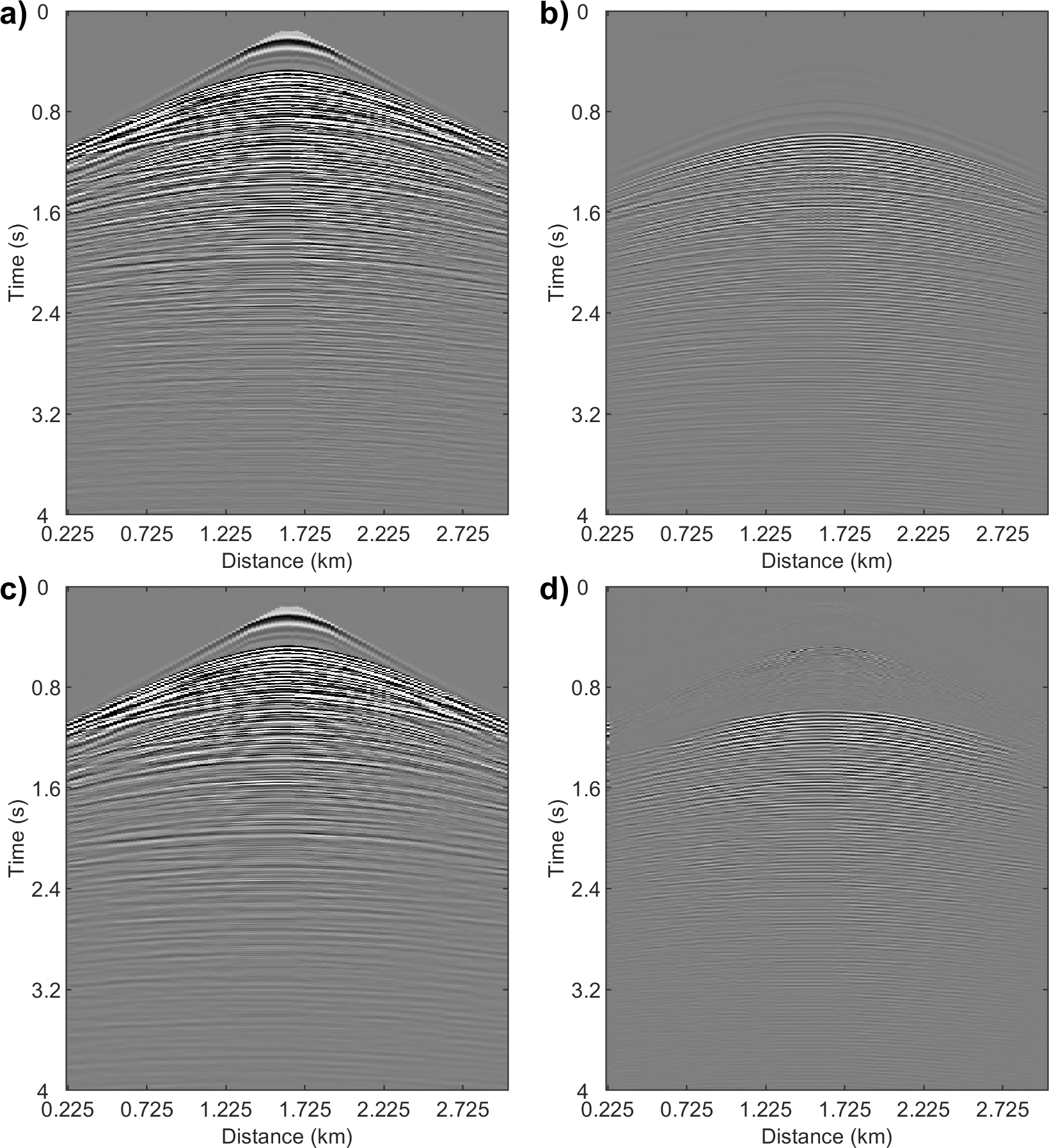}
\caption{(a) An example marine shot gather after preprocessing (near-offset interpolation and reciprocity) with direct arrivals removed. (b) MDC-generated surface multiples, where amplitude is already scaled. (c) Network-predicted result for the gather shown in (a). (d) Residual between the original data (a) and the prediction (c). }
\label{fig11}
\end{figure}

\begin{figure}[htbp]
\centering
\includegraphics[width=0.65\textwidth]{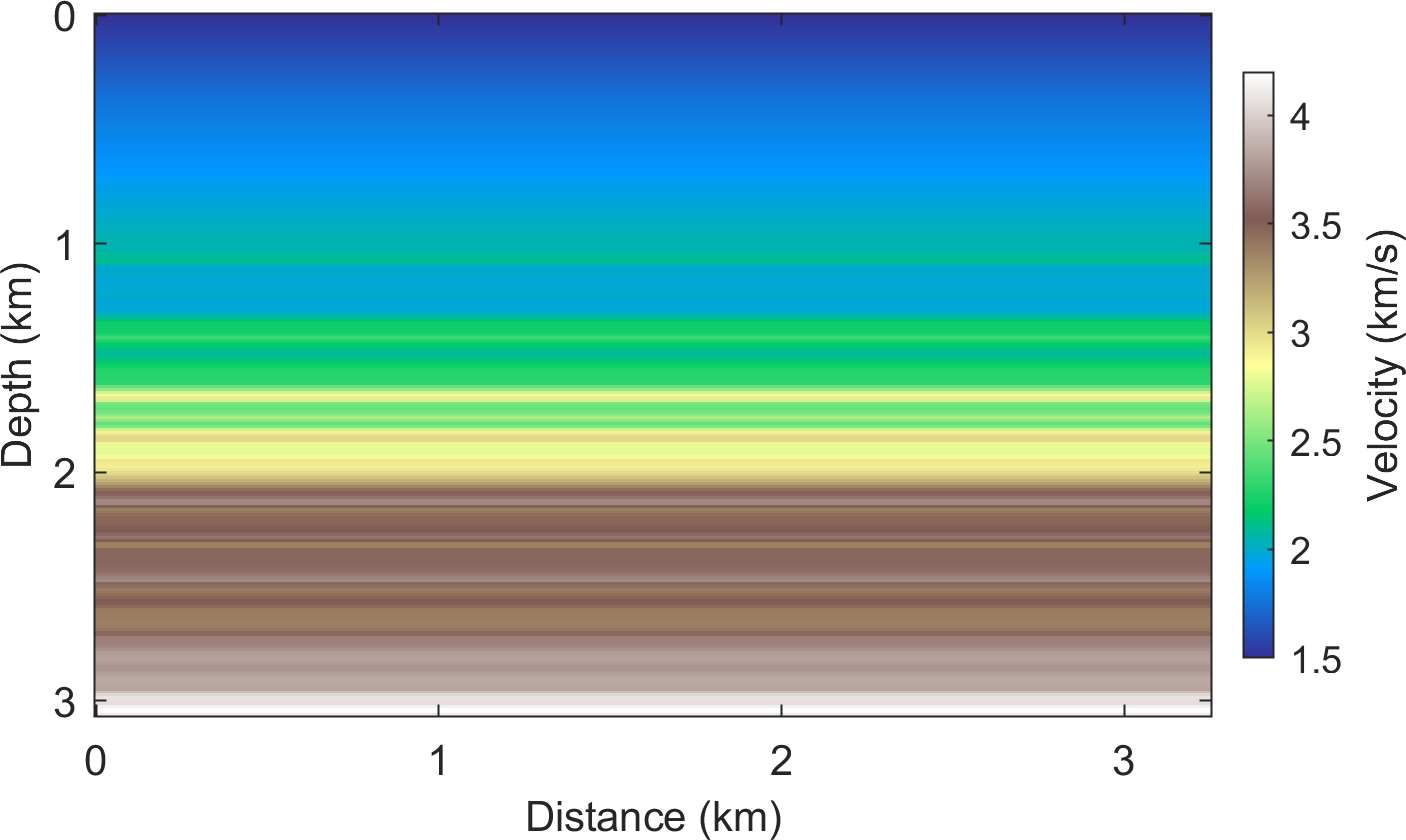}
\caption{Well-based velocity model in the field-data experiment.}
\label{fig12}
\end{figure}

\begin{figure}[htbp]
\centering
\includegraphics[width=0.65\textwidth]{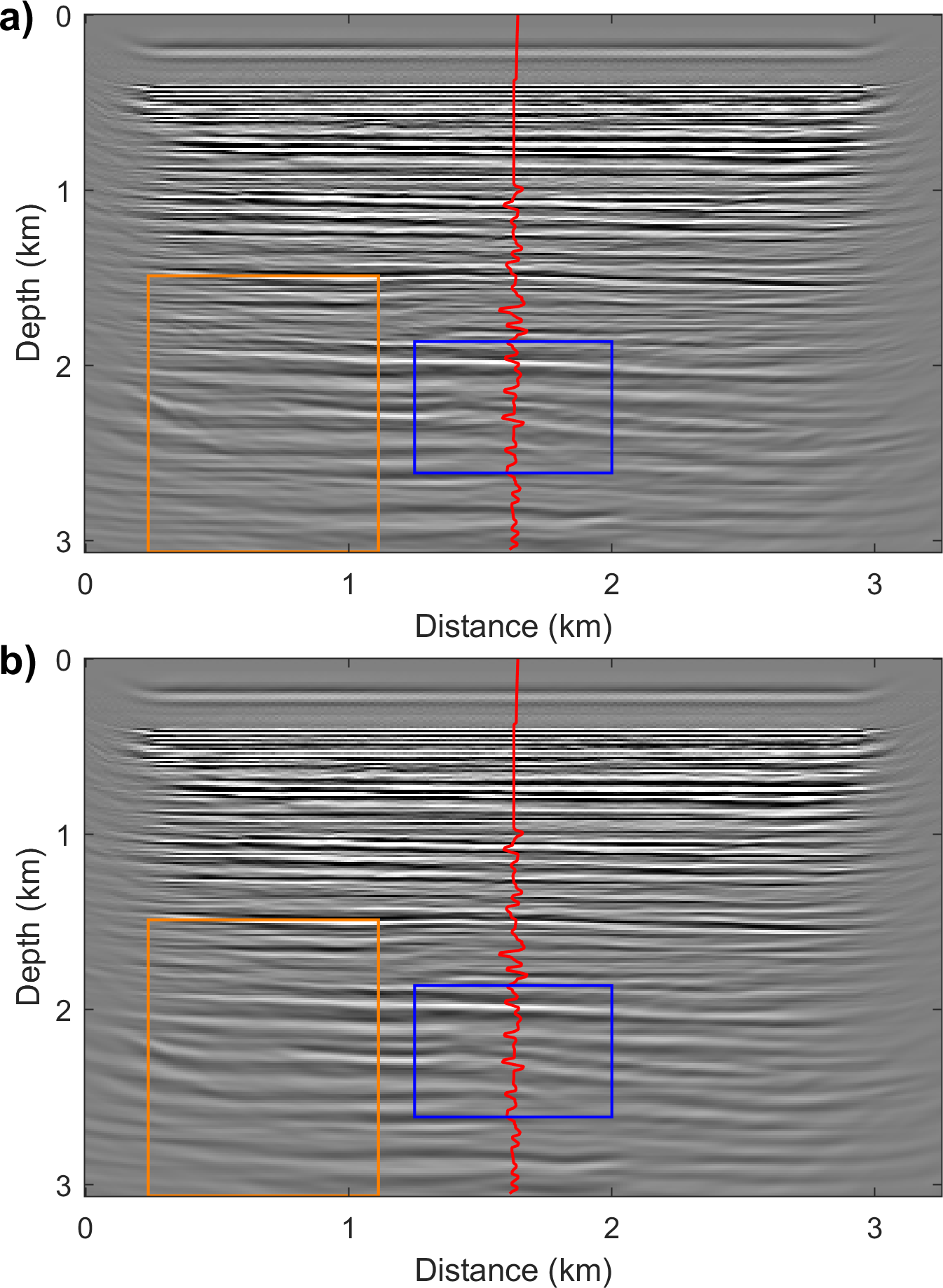}
\caption{Migration results from the marine dataset: (a) Migrated image using the preprocessed data with surface-related multiples. (b) Migrated image using the multiple-suppressed data from our method. The overlaid read lines represent the impedance profile derived from the well.}
\label{fig13}
\end{figure}

\begin{figure}[htbp]
\centering
\includegraphics[width=0.8\textwidth]{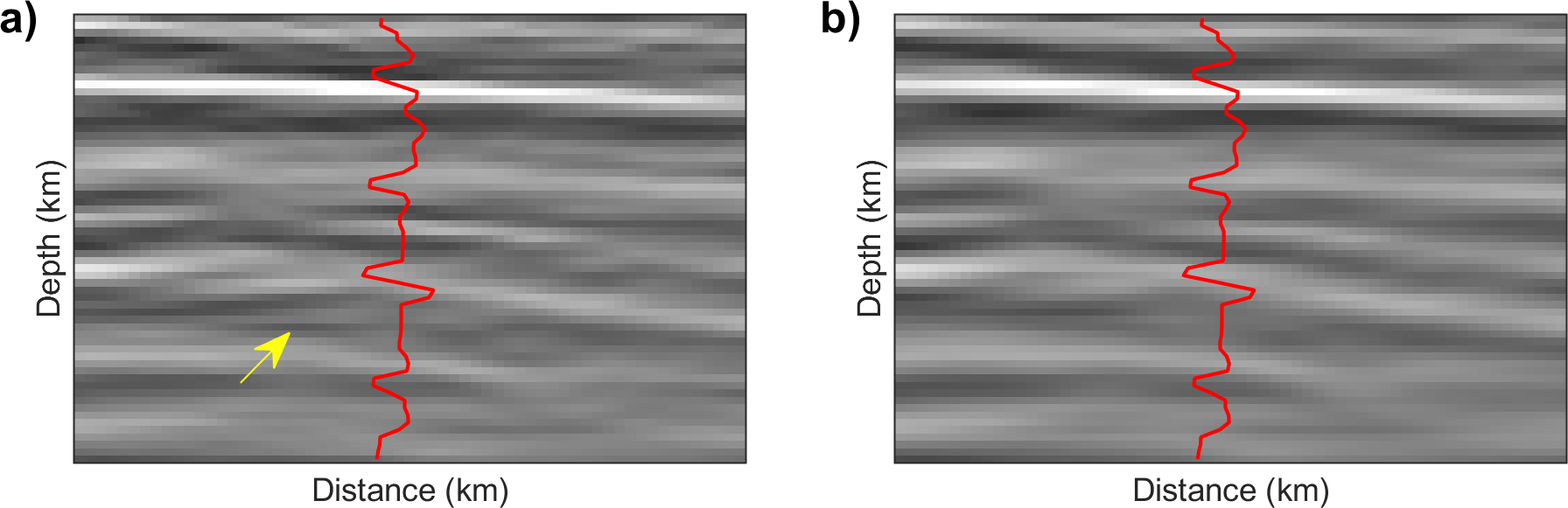}
\caption{Zoomed-in view of the blue-box region in Figure~\ref{fig13}: (a) Original data's migrated image, (b) Multiple-suppressed migrated image.}
\label{fig14}
\end{figure}

\begin{figure}[htbp]
\centering
\includegraphics[width=0.8\textwidth]{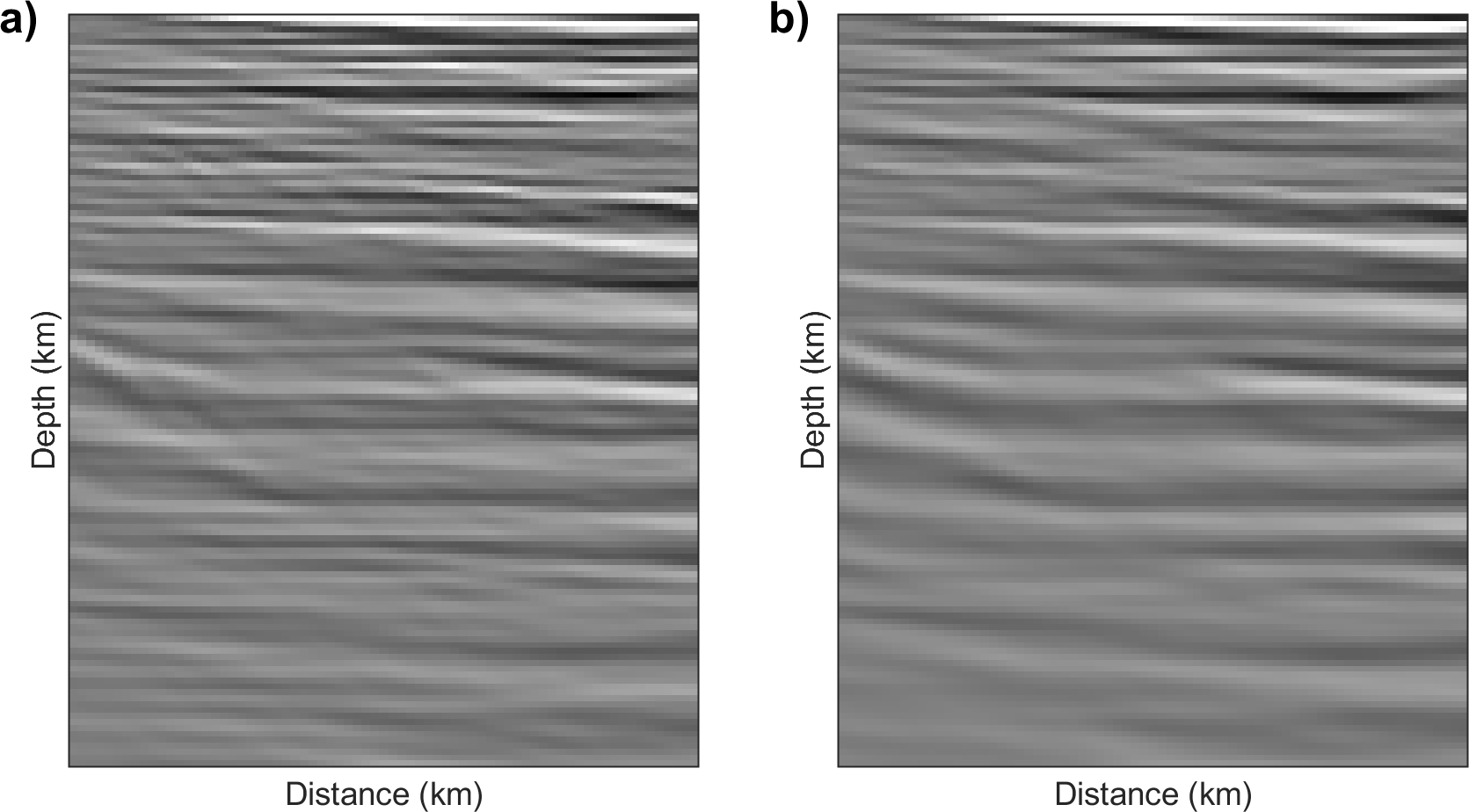}
\caption{Zoomed-in view of the orange-box region in Figure~\ref{fig13}: (a) Original data's migrated image, (b) Multiple-suppressed migrated image.}
\label{fig15}
\end{figure}

\section{\textbf{Field data application}}
Having demonstrated the effectiveness of our method on two synthetic models, we now apply it to an openly available Mobil amplitude variation with offset (AVO) Viking Graben line 12 marine data. Figure~\ref{fig7} illustrates the survey geometry: sources are deployed at a depth of 10 m and placed every 25 m along the survey line, while 120 geophones are located at a depth of 6 m with a receiver spacing of 25 m. Because the receivers trail behind the source, each shot gather lacks near-offset traces for approximately 250 m, resulting in about ten missing near-offset traces per gather and a maximum offset of 3.25 km. 

Figure~\ref{fig8}a shows a representative shot gather from the raw dataset, highlighting the missing near offsets. Since the MDC procedure requires source-receiver collocation, we must reconstruct the near-offset traces. In SRME workflows, Radon- \citep{trad2003interpolation} or Fourier-based interpolation \citep{liu2004minimum} methods are applied when two-sided data are available. Here, however, only single-sided offsets are present. We therefore adopt a simpler approach: we apply a rough normal moveout (NMO) correction to flatten the traces at the near offsets, fill the missing traces by replicating data from the nearest recorded offset, and then remove the NMO correction \citep{curry2010interpolation}. Figures~\ref{fig8}b and \ref{fig8}c illustrate this process. Here, we does not require a precise velocity model. Even a rough velocity profile is sufficient to flatten the near-offset area for interpolation. 

After filling the near offsets, a second challenge remains: the MDC operator requires sources and receivers to overlap, yet the steamer acquisition collects data only on one side of the source. To resolve this constraint, we leverage shot-receiver reciprocity, as schematically shown in Figure~\ref{fig9}. We define a rectangular region (black box) that covers the maximum offset. For a shot $S$ located at $(z_0, x_0)$, if a trace at $(z_0, x_1)$, where $x_1>x_0$, is missing within this region, we retrieve the corresponding trace from the shot whose source is located at $(z_0, x_1)$. By the principle of reciprocity, swapping source and receiver locations is valid for a reciprocal gather. Figure~\ref{fig10} shows three representative shot gathers after applying this procedure, resulting in a two-sided dataset (within the black box) suitable for MDC operation. 

We select shots within the 3.25 km offset range, which contains a well providing velocity and density logs for reference. An example of the preprocessed shot gather, after near-offset interpolation and shot-receiver reciprocity (with direct arrivals removed), is shown in Figure~\ref{fig11}a. We apply MDC to generate synthetic multiples, as displayed in Figure~\ref{fig11}b. Here, we divide the value of the MDC-generated multiples by 10 and then display it within the same range as the original data. These processed gathers and their MDC-generated multiples form our training set. We augment the data by flipping the gathers horizontally and vertically. The batch size is set to 16, and the network is trained for 60 epochs in total, with 50 of those allocated to the warm-up phase. For both warm-up and IDR phases, the range of $\alpha$ is set to $[0.05, 0.12]$. The initial learning rate of $2 \times 10^{-4}$ is reduced to 60\% at the 25th and 50th epochs. Training takes approximately 11 minutes and 40 seconds. 

Figure~\ref{fig11}c shows the network’s prediction for the same shot gather as in Figure~\ref{fig11}a, and Figure~\ref{fig11}d presents the residual. Most of the surface-related multiple energy is suppressed, as evidenced by how well the residual aligns with the MDC-generated multiples. To validate that the suppressed events are indeed surface-related multiples, we perform a migration experiment similar to that in our synthetic examples. Specifically, we perform migration using two datasets: (1) the preprocessed data with surface-related multiples and (2) our predicted data. We construct a velocity model from the well logs, shown in Figure~\ref{fig12}, and apply smoothing before migration. Figures~\ref{fig13}a and \ref{fig13}b display the migrated images using the original and predicted datasets, respectively, overlaid with the impedance profile derived from the well. In contrast to the original data, the image obtained from our predictions is free of some deep sub-horizontal artifacts caused by surface-related multiples. 

To further illustrate these differences, we zoom in on two areas (blue and orange boxes in Figure~\ref{fig13}), shown in Figures~\ref{fig14} and \ref{fig15}. In Figure~\ref{fig14}a, a fake reflection event appears (yellow arrow) where the impedance profile shows no corresponding layer. In Figure~\ref{fig14}b, this artifact is eliminated by our multiple suppression. Likewise, Figure~\ref{fig15} reveals that certain shallow high-resolution events in the original dataset are absent in those from our predicted dataset, consistent with their identification as surface-related multiples. Such high-resolution patterns are commonly induced by multiples traveling in shallow water layers with minimal attenuation. Thus, the field data results confirm that our method effectively attenuates surface-related multiples and yields clearer seismic images. 
\section{\textbf{Discussion}}
Our method shares a fundamental principle with surface-related multiple elimination (SRME): both approaches reconstruct surface-related multiples directly from the recorded seismic data via convolution. In SRME, seismic traces are convolved with themselves to predict the energy of the multiples, and for adaptive subtraction, we need to divide by the source wavelet. This data-driven perspective forms the basis of many traditional multiple-suppression workflows, enabling practitioners to estimate and subtract the unwanted events using the recorded field data. 

Despite this conceptual overlap, our proposed framework diverges in several important ways. First, whereas SRME frequently relies on iterative procedures and adaptive subtraction, we employ a single-pass multi-dimensional convolution (MDC) to generate synthetic multiples at the outset. As a result, SRME can become computationally intensive, especially for large-scale 3D datasets, due to repeated convolutional predictions and adaptive steps. Our approach avoids repeated MDC operations by performing them only once, then leveraging self-supervised learning (SSL) to refine the multiple-suppression performance. This design not only lowers computational overhead but also streamlines the workflow, making it more practical for large volumes of field data. Second, SRME can benefit from or require prior knowledge (e.g., an estimated source wavelet), especially in adaptive subtraction stages. By contrast, our method is an SSL framework and does not depend on any explicit wavelet or velocity-model information, making it more suitable when such information are unavailable or uncertain. A third major difference is in the refinement process: SRME workflows typically involve refining predictions across multiple iterations to account for amplitude and phase mismatches between predicted multiples and real data, where each iteration includes a convolution operation. In our case, the focus is on a two-stage SSL framework, warm-up followed by iterative data refinement (IDR). The network learns progressively from generated pseudo-labels, eliminating the need for multiple-free data and reducing the risk of overfitting to synthetic or model-based assumptions. Therefore, our method maintains the data-driven essence of SRME while introducing an SSL mechanism that requires no labeled data and less reliance on prior knowledge. By combining single-pass MDC with iterative SSL refinement, we provide an efficient and adaptable alternative that can be readily applied to complex geophysical scenarios. 

Nevertheless, A key practical consideration in our method is the choice of the scaling factor \(\alpha\). As illustrated before, an \(\alpha\) value that is too large can lead to over-suppression of primaries, whereas an \(\alpha\) value that is too small may fail to adequately remove surface-related multiples. Currently, determining an appropriate range for \(\alpha\) requires manual inspection of the data to compare the amplitude of MDC-generated multiples with those in the raw observed data. While this approach is generally effective, it introduces a degree of user judgment. Future work will focus on developing an adaptive scheme that automatically learns or updates \(\alpha\) based on data characteristics, potentially further reducing reliance on manual tuning. Such an enhancement would make our framework even more robust and easier to apply across diverse geophysical settings.

\section{\textbf{Conclusions}}
We proposed a self-supervised learning (SSL) framework for suppressing surface-related multiples in seismic data. Our approach employs a single-pass multi-dimensional convolution to generate synthetic multiples from our observed data, which then serve as multiple-augmented inputs in a two-stage SSL procedure: a warm-up phase and an iterative data refinement phase. As a result, our method avoids the need for clean labels and reduces the dependence on prior information such as wavelet estimation or velocity models and, thus, addresses the limitations of traditional and supervised learning methods. Numerical examples on both synthetic and field data demonstrated that our method suppresses surface-related multiples effectively while preserving primary reflections. Migration results confirmed its ability to reduce multiple-induced artifacts and improve imaging quality.

\section{\textbf{Acknowledgment}}
We are very grateful to Matteo Ravasi for sharing his experience on how to perform multi-dimensional convolution operations on streamer data. We also thank Francesco Brandolin for his valuable advice on near-offset interpolation of streamer data. This publication is based on work supported by the King Abdullah University of Science and Technology (KAUST). We also thank the DeepWave sponsors for their support. This work utilized the resources of the Supercomputing Laboratory at King Abdullah University of Science and Technology (KAUST) in Thuwal, Saudi Arabia.
\vspace{0.5cm}
\section{\textbf{Code Availability}}
The data and accompanying codes that support the findings of this study are available at: 
\url{https://github.com/DeepWave-KAUST/SSL-Multiples-Attenuation}. (During the review process, the repository is private. Once the manuscript is accepted, we will make it public.)

\bibliographystyle{unsrtnat}
\bibliography{references}

\end{document}